\documentclass[twocolumn,showpacs,aps,prl,superscriptaddress,floatfix]{revtex4}

\usepackage{graphicx}
\usepackage{dcolumn}
\usepackage{amsmath}
\usepackage{epsfig}

\def\sss{\scriptscriptstyle}
\def\barpd{{\raise.35ex\hbox
{${\sss (}$}}--{\raise.35ex\hbox{${\sss )}$}}}
\def\dbarp{\hbox{$D^{*0}$\kern-1.6em\raise1.5ex\hbox{\barpd}}}
\def\Dstarz  {\ensuremath{D^{*0}}\xspace}
\def\Dstarm  {\ensuremath{D^{*-}}\xspace}
\def\Kstarz  {\ensuremath{K^{*0}}\xspace}

\def\Kstarp  {\ensuremath{K^{*+}}\xspace}

\def\Kpm      {\ensuremath{K^{\pm}}\xspace}
\def\Dz      {\ensuremath{D^0}\xspace}
\def\piz   {\ensuremath{\pi^{0}}\xspace}
\def\pim   {\ensuremath{\pi^{-}}\xspace}
\def\pip   {\ensuremath{\pi^{+}}\xspace}
\def\KS    {\ensuremath{K_{\scriptscriptstyle S}}\xspace}

\def\mes        {\mbox{$m_{\rm ES}$}\xspace}

\def\to         {\ensuremath{\rightarrow}\xspace}

\newcommand{\BaBarYear}       {03}
\newcommand{\BaBarNumber}     {24}
\newcommand{\SLACPubNumber} {10140}

\input babarsym

\begin{document}

{\pagestyle{empty}
\begin{flushleft}
\babar-PUB-\BaBarYear/\BaBarNumber \\
SLAC-PUB-\SLACPubNumber \\
\end{flushleft}
}
\title{
{\large \bf \boldmath
Measurement of the Branching Fraction and Polarization 
for the Decay $B^- \to D^{*0} K^{*-}$ 
}}

%
\author{B.~Aubert}
\author{R.~Barate}
\author{D.~Boutigny}
\author{J.-M.~Gaillard}
\author{A.~Hicheur}
\author{Y.~Karyotakis}
\author{J.~P.~Lees}
\author{P.~Robbe}
\author{V.~Tisserand}
\author{A.~Zghiche}
\affiliation{Laboratoire de Physique des Particules, F-74941 Annecy-le-Vieux, France }
% ...
\collaboration{The \babar\ Collaboration}
\noaffiliation

\date{\today}

\begin{abstract}
\noindent We present a study
of the decay 
$B^- \to D^{*0} K^{*-}$ based on a sample of 
86 million $\FourS\to B\Bbar$ decays collected 
with the \babar\ detector at the
\pep2\ asymmetric-energy \BF\ at SLAC.  We measure the 
branching fraction ${\cal B}(B^- \to D^{*0} K^{*-}$) = $(8.3 \pm 1.1 {\rm
(stat)} \pm 1.0 {\rm (syst)}) \times 10^{-4}$, and the
fraction of longitudinal polarization 
in this decay to be
$\Gamma_L/\Gamma = 0.86 \pm 0.06 {\rm (stat)} \pm 0.03 {\rm (syst)}$.
\end{abstract}


\maketitle

Following the discovery of CP violation in $B$-meson 
decays and the measurement of the angle $\beta$
of the unitarity triangle~\cite{cpv}, focus has turned
towards the 
measurements of the
angles $\alpha$ and $\gamma$.   A precise 
determination of $\gamma$ requires larger samples
of $B$ decays than are currently available, and 
is likely to be based on information from several 
decay modes.  Decays of the type 
$B \to D^{(*)}K^{(*)}$ are expected to play a
leading role in this program~\cite{dk}; among these modes,
those with a $K^{*}$ have distinct advantages 
in some of the proposed methods~\cite{dk2}.
Decay modes into two vector mesons present unique
opportunities due to interference between 
helicity amplitudes.  It has been suggested that
angular analysis of 
$B^- \to \dbarp~K^{*-}$
can yield information on $\gamma$ without
external assumptions~\cite{sinha}.  More generally,
such a study would be sensitive to T-violating
asymmetries that probe physics beyond the Standard Model~\cite{vv}.

The
previously 
available information on 
$B^- \to D^{*0} K^{*-}$  
is based on a sample of 15 events~\cite{cleodstkst}.
Here we report on an improved measurement
of the branching fraction for $B^- \to D^{*0} K^{*-}$,
and on the first measurement of the polarization in this decay.

Results are based on 
86 million $\FourS\to B\Bbar$ decays, 
corresponding to an integrated luminosity of
79 fb$^{-1}$,
collected between 1999 and 2002
with the \babar\ detector at the
\pep2\ 
\BF\ at SLAC~\cite{pep2}.
An additional 9.4~fb$^{-1}$ sample of off-resonance data, recorded 
at $e^+e^-$ center-of-mass (CM) energy 
40~\mev
below the \FourS mass, is used to study
``continuum'' events, $e^+ e^- \to q \bar{q}$
($q=u,d,s,$ or $c$).

The \babar\ detector is described 
elsewhere~\cite{ref:babar}.  Only detector components 
relevant for this analysis are summarized here.  
Trajectories of charged particles are measured in
a five-layer silicon vertex tracker 
(SVT) and a 40-layer drift chamber (DCH) 
in a 1.5-T 
magnetic field.  Charged particles
are identified as pions or kaons using information
from a detector of internally reflected Cherenkov
light, as well as measurements of 
energy loss 
in the
SVT and the DCH. Photons are
detected in a CsI(Tl) calorimeter. 

We reconstruct 
$B^- \to D^{*0} K^{*-}$
in the following modes:
$D^{*0} \to D^0 \pi^0$ and 
$D^0 \gamma$; $D^0 \to K^- \pi^+$, $K^- \pi^+ \pi^0$, and
$K^- \pi^+ \pi^+ \pi^-$; $K^{*-} \to K_S \pi^-$; $K_S \to \pi^+ \pi^-$;
$\pi^0 \to \gamma \gamma$
(charged conjugate decay modes are implied throughout this Letter).
The optimization of the 
event selection was based on studies of 
off-resonance data and 
simulated 
$B\bar{B}$ events.
A key feature of the analysis is the use of a
sample of 
4500 
$B^- \to D^{*0} \pi^-$ events 
to validate several aspects of the 
simulation and the analysis procedure.

We select $K_S$ candidates from pairs of oppositely-charged tracks
that
form an invariant mass within 9 MeV (3$\sigma$)
of the 
known~\cite{ref:PDG} 
$K_S$ mass.
Each $K_S$ candidate is combined with a negatively
charged track, assumed to be a $\pi^-$, to form a
$K^{*-}$ candidate.
We retain
$K^{*-}$ candidates with 
mass within
75 MeV of the 
known 
$K^{*-}$ mass.
The $K_S$ vertex 
must
be displaced by at least 3~mm from
the $K^{*-}$ vertex.
This 
requirement 
rejects
combinatorial 
background and is 96\% efficient for real $K_S$ decays.

Photon candidates are constructed from calorimeter clusters
with lateral profiles consistent with photon showers. 
Neutral-pion candidates are formed from 
pairs of photon candidates with invariant mass between 
115 and 150 MeV.
The $\pi^0$
mass resolution is 6.5 MeV.

We select $D^0$ candidates in the three decay modes listed above.
To reduce backgrounds, tracks from $D^0 \to K^-\pi^+ \pi^0$ and $D^0 \to K^-
\pi^+ \pi^+ \pi^-$ 
must have momenta above 150 \mev.
The kaon candidate track must satisfy a set of 
kaon identification
criteria that provides a rejection factor of about 30 against pions.
The kaon identification efficiency 
averaged over all kinematically allowed momentum and polar angle is 90\%.
For each $D^0 \to K^- \pi^+ \pi^0$ candidate, we calculate the
square of the decay amplitude ($|A|^2$) based on 
the kinematics of the decay products and the
known
properties of the Dalitz plot for this decay~\cite{dalitz}.
We retain candidates 
if $|A|^2$ is greater than 5.5\% of its
maximum possible value.
The efficiency of this requirement is 76\%.
Finally, the measured invariant mass of $D^0$ candidates must be within $2.5
\sigma$ of the $D^0$ mass.

We select $D^{*0}$ candidates by combining $D^0$ candidates with
a $\pi^0$ or photon candidate.
The $\pi^0$ candidate must have
momentum between 70 and 450 \mev in the
CM frame.  
The photon candidate 
must have energy above 100 \mev in the laboratory frame.
We reject photon candidates 
consistent with originating from $\pi^0$ decay
when paired with 
another photon of energy 
above
100 MeV.
We require
the mass difference $\Delta m \equiv m(D^{*0})-m(D^0)$ 
to be between 138.7 and 145.7 
(130.0 and 156.0) MeV  for 
$D^{*0} \to D^0 \pi^0$  
($D^{*0} \to D^0 \gamma$).
The $\Delta m$ resolution 
is 1.1 (6.4) MeV for the $D^0 \pi^0$ ($D^0 \gamma$) mode.

At each stage in the reconstruction chain, the measurement
of the momentum vector of each
intermediate particle is improved by
refitting the 
momenta of its decay products
with kinematic constraints.  
These constraints are based
on the known mass of the intermediate particle
and on the fact that its decay products originate
from a common point in space.

Finally, we select \Bm candidates by combining $D^{*0}$ and
$K^{*-}$ candidates. A \Bm candidate is characterized
by the energy-substituted mass $\mes \equiv \sqrt{(\frac{1}{2} s +
\vec{p}_0\cdot \vec{p}_B)^2/E_0^2 - p_B^2}$ and energy difference $\Delta E 
\equiv E_B^*-\frac{1}{2}\sqrt{s}$, 
where $E$ and $p$ are energy and momentum,
the asterisk
denotes the CM frame, 
the subscripts $0$ and $B$ refer to the
initial \FourS and $B$ candidate, respectively, 
and $s$ is the square of the CM energy.
For signal events we expect $\mes = M_B$
within the experimental resolution of
about 3 MeV,
where $M_B$ is the known $B^-$ mass.

We require $|\Delta E|\le 40$ \mev for \Bm
candidates with a $D^0 \to K^- \pi^+ \pi^0$, and 
$|\Delta E|\le 27.5$ \mev for all other modes.  The 
$\Delta E$ resolution is approximately 19 MeV
in the $K^- \pi^+ \pi^0$ mode and 10 MeV in the
other modes.

To reduce continuum backgrounds, we make use of the ratio of the second
to zeroth order Fox-Wolfram~\cite{foxwolf}
moments ($R_2<0.4$), and the angle
$\theta_T^*$ between the thrust axes of the $B^-$ candidate and the 
remaining 
charged
tracks and 
neutral 
clusters in the event
($|\cos\theta_T^*|<0.85$). We also make requirements
on the polar angle $\theta_B^*$
of the \Bm candidate
($|\cos\theta_B^*|<0.9$), and the energy flow in the rest of the event.
We construct a Fisher discriminant $\cal F$ based on the
energy flow in nine concentric cones around the direction
of the \Bm candidate~\cite{CLEOFisher}. 
We require 
${\cal F}<0.40$ $(0.28)$ for
\Bm candidates with a $D^{*0} \to D^0 \pi^0$ $(D^0\gamma)$.
The energy flow, $\theta_T^*$, and $\theta_B^*$
are all calculated in the CM frame. 
These requirements remove about 80\% of the continuum 
backgrounds and are 79\% (74\%) efficient for signal
in the $D^0 \pi^0$ ($D^0 \gamma$) mode.

In 16\% of the events there is more than one \Bm candidate. In these events we retain the
best candidate based on a $\chi^2$ algorithm that uses the measured
values, known values, and resolutions of the $D^0$ mass and the mass
difference $\Delta m$.

\begin{figure}[hbt]
\centerline{\epsfig{file=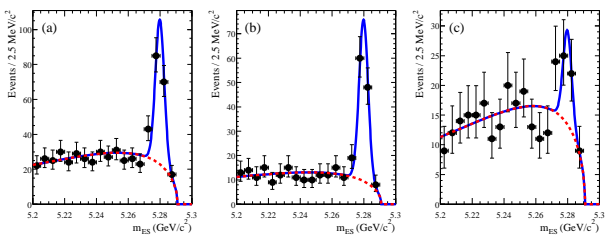, width=3.2in}}
\caption{Distributions of \mes for $B^{-} \to D^{*0}K^{*-}$:
(a) all modes;
(b) $D^{*0} \to D^0 \pi^0$ modes;
(c) $D^{*0} \to D^0 \gamma$ modes.}
\label{fig:mesfit}
\end{figure}

We
extract  
the yield of $B^- \to D^{*0} K^{*-}$ events from a binned
maximum-likelihood fit to the $\mes$ distribution of 
$B^-$
candidates. The signal distribution is parametrized as a
Gaussian and the combinatorial background as a
threshold function~\cite{ARGUS}. The parameters of the
Gaussian are determined from 
the $m_{ES}$ distribution of the $B^- \to D^{*0} \pi^-$
sample.
The total signal yield is $121 \pm 15$
events. The third column of Table \ref{tab:bfbymode2} 
lists the yields for the individual
$D^{*0}/D^0$ 
modes. Figure~\ref{fig:mesfit} shows the $m_{ES}$
distribution of \Bm candidates overlaid with the fit model.

\begin{table*}[tb]
\caption{Summary of the elements of the 
branching fraction calculation.
$N_{m_{ES}}$ is
the yield from the $m_{ES}$ fit;
$N_{pk}$ is the number of peaking background events;
$\epsilon^i_{MC}$ is the
event selection efficiency for the $i$-th mode; 
${\cal B}^i \equiv  {\cal B}_{K^{*-}} \cdot {\cal B}_{K_S} \cdot {\cal B}^i_{D^{*0}} \cdot 
{\cal B}^i_{D^0}$
is the product of branching fractions for the $K^*$, $K_S$, $D^*$, and $D$ decays in 
the $i$-th mode. 
}
\begin{tabular*}{\textwidth}{|l@{\extracolsep{\fill}}|l|c|c|c|c|c|}
\hline
$D^{*0}$ mode & $D^0$ mode & $N_{m_{ES}}$ & $N_{pk}$ & 
\multicolumn{2}{c|}{$\sum{(\epsilon^i_{MC} \times {\cal B}^i)} (\times 10^{-3})$} 
& ${\cal B}(B^- \to D^{*0}K^{*-})(\times 10^{-4})$ \\ \hline 

All & All & $121 \pm 15$ & $6.8 \pm 3.4$ & 
\multicolumn{2}{c|}{$1.6 \pm 0.2$} & $8.3 \pm 1.1 \pm 1.0$ \\ \hline

$D^{*0} \to D^0 \pi^0$ & All & $96 \pm 12$ & $4.8 \pm 2.4$ &
\multicolumn{2}{c|}{$1.0 \pm 0.1$} & $10.2 \pm 1.3 \pm 1.3$ \\ 

$D^{*0} \to D^0 \gamma$ & All & $24 \pm 8$ & $2.0 \pm 1.0$ &
\multicolumn{2}{c|}{$0.6 \pm 0.1$} & $4.4 \pm 1.7 \pm 0.8$ \\ \hline \hline

   &   &    &    & $\epsilon^i_{MC}$ & ${\cal B}^i$ & \\ \hline 

$D^{*0} \to D^0 \pi^0$ & $D^0 \to K^- \pi^+$ & $26 \pm 5$ & $1.7 \pm 0.9$  & $(6.5 \pm 0.6) \%$ &
$(0.54 \pm 0.03) \%$ & $8.0 \pm 1.8 \pm 0.9$ \\  

$D^{*0} \to D^0 \pi^0$ & $D^0 \to K^- \pi^+ \pi^0$ & $39 \pm 8$ & $1.7 \pm 0.9$  & $(2.1 \pm 0.3) \%$ &
$(1.85 \pm 0.15) \%$ & $10.9 \pm 2.4 \pm 1.7$ \\  

$D^{*0} \to D^0 \pi^0$ & $D^0 \to K^- \pi^+ \pi^+ \pi^-$ & $31 \pm 7$ & $1.4 \pm 0.7$  & $(2.9 \pm 0.4) \%$ &
$(1.06 \pm 0.07) \%$ & $11.6 \pm 2.6 \pm 1.6$ \\  

$D^{*0} \to D^0 \gamma$ & $D^0 \to K^- \pi^+$ & $11 \pm 4$ & $0.1 \pm 0.1$  & $(5.7 \pm 0.5) \%$ &
$(0.33 \pm 0.03) \%$ & $6.8 \pm 2.7 \pm 1.0$ \\  

$D^{*0} \to D^0 \gamma$ & $D^0 \to K^- \pi^+ \pi^0$ & $11 \pm 5$ & $1.7 \pm 0.9$  & $(1.9 \pm 0.2) \%$ &
$(1.14 \pm 0.12) \%$ & $5.3 \pm 2.9 \pm 1.0$ \\  

$D^{*0} \to D^0 \gamma$ & $D^0 \to K^- \pi^+ \pi^+ \pi^-$ & $0 \pm 5$ & $0.2 \pm 0.1$  & $(2.5 \pm 0.3) \%$ &
$(0.65 \pm 0.07) \%$ & $-0.2 \pm 3.3 \pm 0.4$ \\  \hline

\end{tabular*}
\label{tab:bfbymode2}
\end{table*}

The 
yield from the $m_{ES}$ fit includes contributions
from ``peaking backgrounds'', which are backgrounds with $m_{ES}$
near $M_B$.
These
backgrounds arise from $B$ decay modes closely
related to the signal mode; e.g., $\bar{B^0} \to D^{*+}K^{*-}$.   
From a 
Monte Carlo simulation
we 
estimate
that 
they
contribute
$6.8 \pm 3.4$ events to the signal yield,
where the 
uncertainty reflects the 
limited knowledge of the branching fractions for 
these modes.

The branching fraction ${\cal B}(B^- \to D^{*0} K^{*-})$ is calculated
from
\begin{equation}
{\cal B} = \frac{N_{m_{ES}} - N_{pk}} {N_{B\bar{B}} 
\cdot {\cal B}_{K^{*-}} \cdot {\cal B}_{K_S} 
\cdot \sum_i{(
\epsilon_{MC}^i \cdot
{\cal B}^{i}_{D^{*0}} \cdot {\cal B}^{i}_{D^{0}}) }}, \nonumber
\end{equation}
 \noindent where $N_{m_{ES}}$ is the event yield from the $m_{ES}$ fit, 
$N_{pk}$ is the
peaking background,
$N_{B \bar{B}} = (85.8 \pm 0.9) \times 10^6$ 
is the number of $B
\bar{B}$ pairs in the data sample, 
${\cal B}_{K^{*-}}$ and 
${\cal B}_{K_S}$ are the branching fractions for 
$K^{*-} \to K_S \pi^-$ and $K_S \to \pi^+ \pi^-$,
$i$ is an index that runs over the six 
$D^{*0}/D^0$ modes considered in this analysis, 
$\epsilon_{MC}^i$ is the event selection efficiency,
and ${\cal B}^{i}_{D^{*0}}$ 
(${\cal B}^{i}_{D^{0}}$)
is the $D^{*0}$ ($D^{0}$) branching fraction
for the $i$-th mode.
This calculation assumes ${\cal B}(\FourS \to B^+ B^-) = 
{\cal B}(\FourS \to B^0 \bar{B^0})$. 
The Monte Carlo efficiency
determination 
uses the value of the polarization reported in this Letter.

The inputs to the branching fraction calculation
are summarized in Table~\ref{tab:bfbymode2}.  
Combining the six $D^{*0}/D^0$ modes, we measure
a branching fraction
\begin{equation}
{\cal B}(B^- \to D^{*0} K^{*-}) = 
(8.3 \pm 1.1 ({\rm stat}) \pm 1.0 ({\rm syst})) \times 10^{-4}. \nonumber
\end{equation}

The reconstruction efficiencies for photons and charged 
tracks 
are understood at the level of 2.5\%
per photon and 0.8\% per track, based on studies of a variety
of control samples.  These are the dominant systematic 
uncertainties in the determination of ${\cal B}$.   
The efficiencies of many
of the analysis requirements are measured in the
large 
$B^- \to D^{*0} \pi^-$ 
control 
sample. 
The uncertainties on ${\cal B}$ are listed in
Table \ref{tab:bfsys}.

We also compute 
the branching fraction 
${\cal B}(B^- \to D^{*0} K^{*-})$
using only events from
the individual $D^{*0}/D^0$ modes (see Table \ref{tab:bfbymode2}). 
The branching fractions measured in the 
$D^{*0} \to D^0 \gamma$ modes are somewhat lower than
those measured in the $D^{*0} \to D^0 \pi^0$ modes.
However, in the 
$B^- \to D^{*0} \pi^-$ 
sample, where the $D^{*0}$ is reconstructed with identical techniques, we find 
agreement between data and expectations
for the relative yields of events in all six
modes.
Thus, we ascribe the difference in the measured 
branching fractions between the modes listed
in Table~\ref{tab:bfbymode2} to statistical fluctuations.
\begin{table}[hbt]
\caption{Uncertainties on ${\cal B}(B^- \to D^{*0} K^{*-})$.}
\begin{tabular}{|l|c|}
\hline
Source                   &    Uncertainty \\ \hline
Statistical               &    13.1\%            \\ \hline
$\pi^0$ and $\gamma$ efficiency  &     6.0\%             \\
Tracking efficiency      &     4.5\%             \\
\mes fitting assumptions &     3.8\%             \\
Event selection criteria &     3.8\%             \\
$D^{*0}$ and $D^0$ branching fractions  &     3.2\%             \\
Peaking background estimates     &     3.0\%             \\
Kaon identification efficiency        &     2.0\%             \\
$K_S$ efficiency       &     1.9\%             \\
Polarization uncertainty             &     1.8\%             \\
Monte Carlo statistics   &     1.7\%            \\
$N_{B \bar{B}}$            &     1.1\%             \\
\hline
Total Systematics        &    11.7\%             \\
\hline
\end{tabular}
\label{tab:bfsys}
\end{table}

The angular distributions for the decay chains 
$B^- \rightarrow D^{*0}K^{*-}$ followed by
$D^{*0} \rightarrow D^0 \pi^0$ or 
$D^0 \gamma$ are expressed in terms of
three 
amplitudes $H_0$ (longitudinal), $H_+$, and 
$H_-$(transverse), and three angles, $\theta_{D}$, 
$\theta_{K}$ and $\chi$~\cite{poltheory}.
The angle $\theta_{D}$ ($\theta_K$) is
the 
angle of the $D^0$ ($K_S$)
with respect to the $B^-$ direction in the $D^{*0}$ 
($K^{*-}$) rest frame;
$\chi$ is the angle between the decay planes of the $D^{*0}$ and the
$K^{*-}$ in the $B^-$ rest frame.  
The experimental acceptance is nearly independent of $\chi$.
Integrating over $\chi$,
the angular distributions reduce to
\begin{equation}
\rm
\begin{array}{lcl}
\frac{d^2\Gamma}{d\cos\theta_Dd\cos\theta_{K}} & \propto &
4|H_0|^2 \cos^2\theta_{D} \cos^2\theta_{K} \\
& + &(|H_{+}|^2 + |H_{-}|^2) \sin^2\theta_{D} \sin^2\theta_{K}, \\
\frac{d^2\Gamma}{d\cos\theta_{D} d\cos\theta_{K}} & \propto &
4|H_0|^2 \sin^2\theta_{D} \cos^2\theta_{K} \\
& + & (|H_{+}|^2 + |H_{-}|^2)(1+\cos^2\theta_{D}) \sin^2\theta_{K}
\end{array} \nonumber
\end{equation}
\noindent for $D^{*0} \rightarrow D^0 \pi^0$ and
$D^{*0} \rightarrow D^0 \gamma$, respectively.

The longitudinal polarization fraction $\Gamma_L/\Gamma$, given by
\begin{equation}
\frac{\Gamma_L}{\Gamma} = \frac{|H_0|^2}{|H_0|^2+|H_{+}|^2 + |H_{-}|^2}, \nonumber
\end{equation}
\noindent 
is extracted
from
an unbinned maximum-likelihood fit to the
$(\theta_{D},\theta_{K})$ distribution for events
with $m_{ES}>5.27$ GeV. The data distribution ($D$) 
is fit to the sum of distributions for
longitudinally
($L$) and transversely ($T$) polarized signal
events,
and combinatorial background events ($C$):
\begin{equation}
D(\theta_{D},\theta_{K}) = a \cdot L(\theta_{D},\theta_{K})
                         + b \cdot T(\theta_{D},\theta_{K}) 
                         + c \cdot C(\theta_{D},\theta_{K}).
\nonumber
\end{equation}

\noindent Here $c$ is the fraction of combinatorial
background determined from the $m_{ES}$ yield fit,
and $b = 1 - a - c$.
Thus, $a$ is the only free parameter in the fit.

The distributions of $L$ and $T$ are obtained from 
simulations of transverse and longitudinal decays,
including detector acceptance effects.
The distribution of $C$ is estimated from data 
candidates in a sideband of $\mes$ ($5.20< \mes <5.27$ GeV). 
We exclude from the fit ($\theta_D,\theta_K$)
regions where the efficiency changes rapidly: 
$\cos\theta_K < -0.9$ and, in the $D^0 \gamma$
mode, $\cos\theta_D > 0.85$.

\begin{figure}[hbt]
\centerline{
\epsfig{file=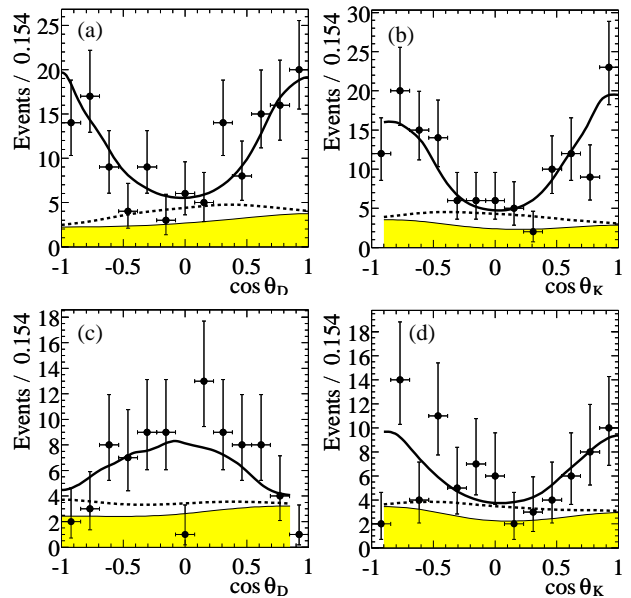, height=8cm}
}
\caption{Distributions of (a) $\cos\theta_{D}$ and
(b) $\cos\theta_{K}$
for $D^{*0} \to D^0 \pi^0$.
Distributions of (c) $\cos\theta_{D}$ and
(d) $\cos\theta_{K}$
for $D^{*0} \to D^0 \gamma$.
The solid line represents
the full fit model, the dashed line represents the transverse
component, and the shaded region represents the combinatorial background
component. 
}
\label{fig:pol}
\end{figure}

We find longitudinal polarization fractions $\Gamma_L/\Gamma =
0.87 \pm 0.07 ({\rm stat})\pm 0.03 ({\rm syst})$ and
$0.80 \pm 0.14 ({\rm stat}) \pm 0.04 ({\rm syst})$
from fits to the $D^{*0} \to D^0 \pi^0$ 
and $D^{*0} \to D^0 \gamma$ samples, respectively.
Figure~\ref{fig:pol} shows 
projections of the 
$(\theta_{D},\theta_{K})$
distributions for the event sample.

Combining the results from the two $D^{*0}$ modes, we find 
$\Gamma_L/\Gamma = 0.86 \pm 0.06 ({\rm stat}) \pm 0.03 ({\rm syst})$.
The systematic uncertainty 
reflects the accuracy of the simulation ($\pm$ 0.017), 
the uncertainty on the fraction $c$ ($\pm$ 0.017),
the finite statistics of the simulation ($\pm$ 0.010),
the uncertainties related to the fit
assumptions ($\pm$ 0.010), and the uncertainty 
due to the assumption that the acceptance is independent of $\chi$ ($\pm$ 0.004).
As a consistency check, we fit the $\theta_{D}$
distribution in the 
$B^- \to D^{*0} \pi^-$
sample.  We find $\Gamma_L/\Gamma = 1.00 \pm 0.01$,
in agreement with the expectation $\Gamma_L/\Gamma = 1$  
from angular momentum conservation.

In summary, we have measured the branching fraction 
for $B^- \to D^{*0} K^{*-}$ to be
${\cal B} = (8.3 \pm 1.1 ({\rm stat}) \pm 1.0 ({\rm syst})) 
\times 10^{-4}$. Our
measurement 
is 
a factor of 2.5 more precise than
the previous result.  It is in agreement with predictions 
based on the measured $B^- \to D^{*0} \rho^-$ 
branching fraction~\cite{dstrho},
and the value of the Cabibbo angle.
We have also measured 
the
longitudinal polarization fraction in this decay to be $\Gamma_L /
\Gamma = 0.86 \pm 0.06 ({\rm stat}) \pm 0.03 ({\rm syst})$.  This
last result is 
consistent with
expectations~\cite{neubert}
based
on factorization, Heavy Quark Effective Theory, and the measurement
of semileptonic $B$-decay form factors, assuming
that the 
external spectator
amplitude 
($b \to c W^{*-}$; $W^{*-} \to K^{*-}$)
dominates in $B^- \to D^{*0} K^{*-}$.

We are grateful for the excellent luminosity and machine conditions
provided by our \pep2\ colleagues, 
and for the substantial dedicated effort from
the computing organizations that support \babar.
The collaborating institutions wish to thank 
SLAC for its support and kind hospitality. 
This work is supported by
DOE
and NSF (USA),
NSERC (Canada),
IHEP (China),
CEA and
CNRS-IN2P3
(France),
BMBF and DFG
(Germany),
INFN (Italy),
FOM (The Netherlands),
NFR (Norway),
MIST (Russia), and
PPARC (United Kingdom). 
Individuals have received support from the 
A.~P.~Sloan Foundation, 
Research Corporation,
and Alexander von Humboldt Foundation.

\end{document}